\documentclass[prl,twocolumn,amsmath,amssymb,aps,superscriptaddress,flushbottom]{revtex4-2}

\newcommand{\DVred}[1]{\textcolor{black}{#1}}

\newcommand{\DVblue}[1]{\textcolor{black}{#1}}

\usepackage[utf8]{inputenc}
\usepackage[T1]{fontenc}
\usepackage{graphicx}
\usepackage{float}
\usepackage{color}
\usepackage{xr}
\usepackage{bm}
\usepackage{ulem}
\usepackage{enumerate}
\usepackage{hhline}
\usepackage{mathrsfs}

\newcommand{\ktr}{k_{\mathrm{tr}}}
\newcommand{\ntr}{n_{\mathrm{tr}}}
\newcommand{\kn}{k_{n}}
\newcommand{\KVK}[1]{\textcolor{black}{#1}}
\newcommand{\un}{u_{n}}
\newcommand{\nln}{\Phi_{n}}
\newcommand{\fn}{f_{n}}
\newcommand{\bel}{\begin{equation}\label}
\newcommand{\ee}{\end{equation}}
\newcommand{\beq}{\begin{eqnarray}\label} 
\newcommand{\eeq}{\end{eqnarray}}

\begin{document}
 \title{Turbulent cascade arrests and the formation of intermediate-scale
condensates
 }
\author{Kolluru Venkata Kiran}
\affiliation{Centre for Condensed Matter Theory, Department of Physics, Indian Institute of Science, Bangalore, 560012, India}

\author{Dario Vincenzi}
\affiliation{Universit\'e C\^ote d'Azur, CNRS, LJAD, 06100 Nice, France}

\author{Rahul Pandit}
\affiliation{Centre for Condensed Matter Theory, Department of Physics, Indian Institute of Science, Bangalore, 560012, India}

\date{\today}

\begin{abstract}
Energy cascades lie at the heart of the dynamics of turbulent flows.
In a recent study of turbulence in fluids with odd-viscosity [de Wit \textit{et al.}, Nature \textbf{627}, 515 (2024)], the two-dimensionalization of the flow at small scales leads to the arrest of the energy cascade and selection of an intermediate scale, between the forcing and the viscous scales. \KVK{To demonstrate the generality of the phenomenon and its existence for a wide class of turbulent systems,}
we study a shell model that is carefully constructed to have three-dimensional turbulent dynamics at small wavenumbers and two-dimensional turbulent dynamics at large wavenumbers.  The large scale separation that we can achieve in our shell model allows us to examine clearly the interplay between these dynamics,
which leads to an arrest of the energy cascade at a transitional wavenumber and an associated accumulation of energy at the same scale. Such pile-up of energy around the transitional wavenumber is reminiscent of the formation of condensates in two-dimensional turbulence, \textit{but, in contrast, it occurs at intermediate wavenumbers instead of the smallest wavenumber}. 
\end{abstract}

\maketitle


\paragraph{Introduction.} Cascade processes are at the origin of the multiscale nature of turbulent flows. The best-known example is Richardson's cascade of kinetic energy in three dimensions (3D)~\cite{frisch1995}. Energy is injected into a characteristic wavenumber $k_{f}$ by an external force that drives the flow. On average, nonlinear interactions transfer the injected energy to larger wavenumbers $k$ without dissipation. This process persists up to the Kolmogorov wavenumber, beyond which viscous dissipation dominates. The continuous interplay between energy injection and viscous dissipation leads to a non-equilibrium statistically stationary state. The direct energy cascade is at the heart of 3D homogeneous isotropic turbulence~\cite{frisch1995}. 
In two-dimensional (2D) turbulence, the direction of the energy cascade is reversed, and on average energy flows from $k_f$ to smaller values of $k$; furthermore,
the inverse cascade of energy coexists with a direct cascade of enstrophy~\cite{be12,review-pandit}. In general, turbulent cascades of inviscid invariants, and in particular their directions (whether direct or inverse), are affected by phenomena such as rotation, stratification, spatial confinement, and selective suppression of Fourier modes of the velocity~\cite{fppr2012,ab18,verma2019,pouquet2019helicity}.

Dissipative mechanisms inherent to the system, typically viscous dissipation at large $k$ or frictional dissipation at small $k$, naturally arrest or suppress a turbulent cascade. \KVK{However, other complex mechanisms that induce such suppression have been identified in geostrophic~\cite{lvf1999,ssd2007,knobloch22}, magnetohydrodynamic~\cite{mdps2021,benavides22}, and bacterial turbulence~\cite{linkmann2019phase,pkp23}.}
\KVK{In particular, de Wit \textit{et al}.~\cite{wfktv23} recently studied the 3D Navier--Stokes equations (NSE) with an odd-viscosity term that leads to a quasi-two-dimensionalization of the velocity field at large $k$.} If the forcing acts at small $k$, the direct energy cascade is arrested at an intermediate wavenumber $k_c$. Since odd viscous terms are not dissipative, the arrest of the cascade results in an accumulation of energy around $k_c$ and thence an emergence of flow structures of size $k_c^{-1}$. If the flow is forced at large $k$, this model displays an inverse energy cascade, which is arrested by the 3D-type behavior of the flow at small $k$ and is accompanied by the formation of structures of intermediate sizes.

\KVK{We demonstrate that such intermediate-scale cascade arrests are not restricted to odd-viscous fluids but are, in fact, signatures of any turbulent systems where 3D-type dynamics, at small $k$, coexist with 2D-type dynamics, at large $k$. To this end}, we construct a shell model that can display the desired small- and large-$k$ dynamics. Shell models are a class of hydrodynamical equations, \KVK{which retain the essential features of NSE in Fourier space} and offer insights into energy-transfer mechanisms in fully developed turbulence~\cite{frisch1995,bjpv98,basu1998multiscaling,b03,plunian2013shell}. The interactions between the Fourier modes of the velocity are restricted; thus, numerical simulations of shell models reach larger scale separations and better statistical convergence than those presently achievable with  direct numerical simulations of the NSE~\cite{de2024extreme}. The tractability of shell models makes them invaluable for studying turbulence. New concepts in turbulence theory that have been recently developed by using shell models include hidden scale invariance~\cite{m21}, subgrid closures~\cite{bmp17}, stirring strategies for optimizing mixing~\cite{md18}, and the application of avalanche dynamics in amorphous materials in the analysis of the temporal behavior of the kinetic energy~\cite{bctt22}.
We \KVK{modify} the strategy that was used by Boffetta \textit{et al}.~\cite{bdlm11} for a shell-model study of thin fluid layers. By allowing the coefficients of the shell model to depend suitably on $k$, the model captured the split-energy-cascade, in quasi-2D turbulent flows, with direct and inverse components~\cite{cmv10,musacchio2017split}. We consider scale-dependent coefficients but select them so as to obtain a shell model that is 3D-like at small wavenumbers and 2D-like at large wavenumbers. The transition between the two cascading regions occurs at a wavenumber $k_{\mathrm{tr}}$; the forcing is localized at the wavenumber $k_{f}$. We find that, irrespective of the ratio of $k_{\rm tr}/k_{f}$, the reciprocity between the small- and large-wavenumber dynamics results in the arrest of the energy cascade (be it forward or inverse). However, the system dynamics differs depending on whether  $\ktr/k_f$ is smaller or greater than unity. The main consequence of the arrest of the cascade is a strong build-up of energy close to $\ktr$. We show that a statistically stationary state \KVK{and the formation of condensates} is nevertheless possible because of an increase in viscous dissipation near $k_{\rm tr}$. 
\paragraph{Model.}We consider the SABRA shell model~\cite{l1998improved,gilbert02}
\bel{shelleq}
\frac{d \un}{dt}=\mathrm{i}\nln-(\mu \kn^{-2}+\nu \kn^{2})\un+\fn, ~~1 \le n\le N,
\ee
where $\kn=k_{0}\lambda^{n}$,
$\mu$ and $\nu$ are the hyper-friction and viscosity parameters, respectively, and the external forcing, $\fn=\epsilon_{f}(1+\mathrm{i})\delta_{n,n_{f}}/2u^{*}_{n_{f}}$, injects energy at a constant rate $\epsilon_{f}$ into the shell $n_{f}$.
The nonlinear term is 
\beq{nlin}
\nln&=&a_{n}k_{n+1}u_{n+2}u^{*}_{n+1}+b_{n}k_{n}u_{n+1}u^{*}_{n-1} \nonumber \\
&-&c_{n}k_{n-1}u_{n-2}u_{n-1},
\eeq
where $a_{n}$, $b_{n}$, $c_{n}$ are real and the $\ast$ denotes complex conjugation.
In addition, $u_{-1}=u_{0}=u_{N+1}=u_{N+2}=0$. 
In the inviscid ($\mu=0$, $\nu=0$) and unforced ($\fn=0$) case, the total energy $E(t)=\frac{1}{2}\sum^{N}_{n=1}|\un(t)|^{2}$ is conserved provided $a_{n-1}+b_{n}+c_{n+1}=0$.
In the original SABRA model~\cite{l1998improved}, $a_{n}=a$, $b_{n}=b$, and $c_{n}=c$, and $H(t)=\frac{1}{2} \sum^{N}_{n=1}\left(a/c\right)^n|\un(t)|^{2}$ is a second invariant quantity. Different regimes are observed depending on the ratio $c/a$. 
Here, it is sufficient to recall that there is a regime of direct energy cascade, which mimics 3D turbulence, for $-1<c/a<0$ \cite{l1998improved}. In this regime, $H$ does not have a definite sign and it can be regarded as a generalized helicity. By contrast, $H$ is positive for $0<c/a<1$, and in the subrange $\lambda^{-2/3}<c/a<1$ there is a 2D-turbulence-like regime, with a simultaneous inverse cascade of $E$ and a direct cascade of $H$. \DVblue{In this case,} $H$ plays the role of a generalized enstrophy~\cite{gilbert02}. 

\begin{table}[t!]
		\centering
		\begin{tabular}{|c| c| c| c|c|c| c|c|}
			\hline
			Run & $N$ &  $n_{f}$& $\ntr$& $\nu$ & $\mu$ & $\delta t$ &$\ktr/k_{\nu}$
			\\
			\hline
			A0& $28$& $1$&$\infty$&$5\times10^{-7}$&0&$1\times10^{-4}$&$\infty$\\
			\hline
         	A1& $28$& $1$&$25$&$5\times10^{-7}$&$0$&$1\times10^{-4}$&$1.48\times 10^{2}$\\
                \hline
                A2& $28$& $1$&$20$&$5\times10^{-7}$&$0$&$1\times10^{-4}$&$4.63\times 10^{0}$\\
                \hline
                A3& $28$& $1$&$18$&$5\times10^{-7}$&$0$&$1\times10^{-4}$&$1.16\times 10^{0}$\\
                \hline
                A4& $28$& $1$&$15$&$5\times10^{-7}$&$0$&$1\times10^{-4}$&$1.45\times 10^{-1}$\\
                \hline
                A5& $28$& $1$&$12$&$5\times10^{-7}$&$0$&$1\times10^{-4}$&$1.18\times 10^{-2}$\\
                \hline
                A6& $28$& $1$&$10$&$5\times10^{-7}$&$0$&$5\times10^{-5}$&$4.50\times 10^{-3}$\\
                \hhline{|=|=|=|=|=|=|=|=|}
                \hline
                B1& $28$& $1$&$15$&$1\times10^{-7}$&$0$&$1\times10^{-4}$&$4.33\times 10^{-2}$\\
                \hline
                B2& $28$& $1$&$15$&$1\times10^{-8}$&$0$&$1\times10^{-5}$&$7.77\times 10^{-3}$\\
                \hline
                B3& $28$& $1$&$15$&$5\times10^{-9}$&$0$&$1\times10^{-5}$&$4.60\times 10^{-3}$\\
            \hhline{|=|=|=|=|=|=|=|=|}
                \hline
            C1& $34$& $3$&$5$&$1\times10^{-10}$&$1\times10^{-3}$&$1\times10^{-4}$&$2.38\times 10^{-7}$\\
                \hline
           C2& $34$& $3$&$7$&$1\times10^{-10}$&$1\times10^{-3}$&$1\times10^{-4}$&$9.51\times 10^{-7}$\\
                \hline
           C3& $34$& $3$&$9$&$1\times10^{-10}$&$1\times10^{-3}$&$1\times10^{-4}$&$3.80\times 10^{-7}$\\
     
                            \hhline{|=|=|=|=|=|=|=|=|}
                \hline

                D1& $34$& $25$&$0$&$1\times10^{-10}$&$5\times10^{-4}$&$1\times10^{-5}$&0\\
                
                \hline
                D2& $34$& $25$&$15$&$1\times10^{-10}$&$5\times10^{-4}$&$1\times10^{-6}$&$2.44\times 10^{-4}$\\
                \hline
                D3& $34$& $25$&$17$&$1\times10^{-10}$&$5\times10^{-4}$&$1\times10^{-6}$&$9.74\times 10^{-4}$\\
                \hline
                D4& $34$& $25$&$18$&$1\times10^{-10}$&$5\times10^{-4}$&$1\times10^{-6}$&$1.90\times 10^{-3}$\\
                \hline
                D5& $34$& $25$&$20$&$1\times10^{-10}$&$5\times10^{-4}$&$1\times10^{-6}$&$7.80\times 10^{-3}$\\
                \hline
		\end{tabular}
		\caption{\label{tab:widgets} Parameters of the shell-model simulations. In addition, $\lambda=2$, $k_0=1/16$, and $\epsilon_{f}=5\times10^{-3}$ for all runs. 
  \DVred{The initial condition is $u_n=k_n^{1/2}e^{\mathrm{i}\theta_n}$ for $n=1,2$ and $u_n=k_n^{1/2}e^{-k_n^2}e^{\mathrm{i}\theta_n}$ for $3\leq n\leq N $, where $\theta_n$ is a random variable distributed uniformly between 0 and $2\pi$.}
}
\label{tab:parameters}
	\end{table}

To introduce a shell-model analog of the domain aspect ratio in a study of quasi-2D fluid turbulence,
Ref.~\cite{bdlm11} considered a version of the SABRA model with $n$-dependent coefficients: 
$\{a_{n},b_{n},c_{n}\}$ were chosen to generate an inverse energy cascade for $k_{n}<k_{h}$ and a direct energy cascade for  $k_{n}>k_{h}$, with $k^{-1}_{h}$ representing the depth of a fluid layer.

We consider a shell model with $n$-dependent coefficients \textit{but reverse the directions of the energy cascades} by taking
\begin{align}{\label{model}}
&a_{n}=1, ~ b_{n}=-0.5,  ~c_{n}=-0.5,~~ 1 \le n< \ntr\,,  \\
&a_{n}=1, ~ b_{n}= -1.7, ~c_{n}=-0.5, ~~ n=\ntr\,, \\
&a_{n}=1,  ~b_{n}= -1.7,  ~c_{n}=0.7,~~\ntr < n\le N\,. 
\end{align}
Clearly, for modes $n<\ntr$ ($n>\ntr$) the coefficients $\{a_{n},b_{n},c_{n}\}$ lead to a 3D (2D) turbulent-like regime. Note that $c_{n}$ changes value at $n_{\mathrm{tr}+1}$ to respect energy conservation in the inviscid limit. 

We investigate the interplay between the small- and large-$k_n$ modes in this model for the cases (a) $\ktr/k_{f} > 1$ and (b) $\ktr/k_{f} < 1$, with 
$\ktr=k_0\lambda^{\ntr}$ the transitional wavenumber. We integrate Eq.~\eqref{shelleq} by using an Adams-Bashforth scheme~\cite{pisarenko93}.

\paragraph{(a) Small-wavenumber forcing.} We first consider the case $k_{f}<\ktr$ (Table~\ref{tab:parameters}, runs~A and~B, in which $\mu=0$).
In the limit $\ktr/k_f\to\infty$, we recover the SABRA model with constant 3D-like coefficients. \DVblue{Therefore,} in this limit, our model
displays a direct cascade of $E$~\cite{l1998improved}, and the energy flux [$\langle \cdot \rangle$ is the
time average]
\bel{Eflux}
\Pi_E(\kn)=\bigg\langle\sum^{N}_{j=n}\Re \big\{\mathrm{i}\Phi_{j}u^{*}_{j}\big\}\bigg\rangle
\ee
is constant and equal to $\epsilon_f$ for $k_{f}\ll \kn \ll k_{\nu}$, 
where $k_\nu=(\epsilon_f/\nu^3)^{1/4}$ is the Kolmogorov wavenumber. In the same range, the energy spectrum $\mathcal{E}(\kn)=\big\langle\vert \un \vert^{2}/\kn\big\rangle $ shows a scaling range that is consistent with $k^{-5/3}_{n}$, the Kolmogorov (1941) form~\cite{frisch1995}. For $\kn \ll k_{f}$, $\mathcal{E}(\kn)\sim k^{-1}_{n}$, which indicates energy equipartition~\cite{ab18}.
\begin{figure}[!]
 
      \includegraphics[width=\columnwidth]{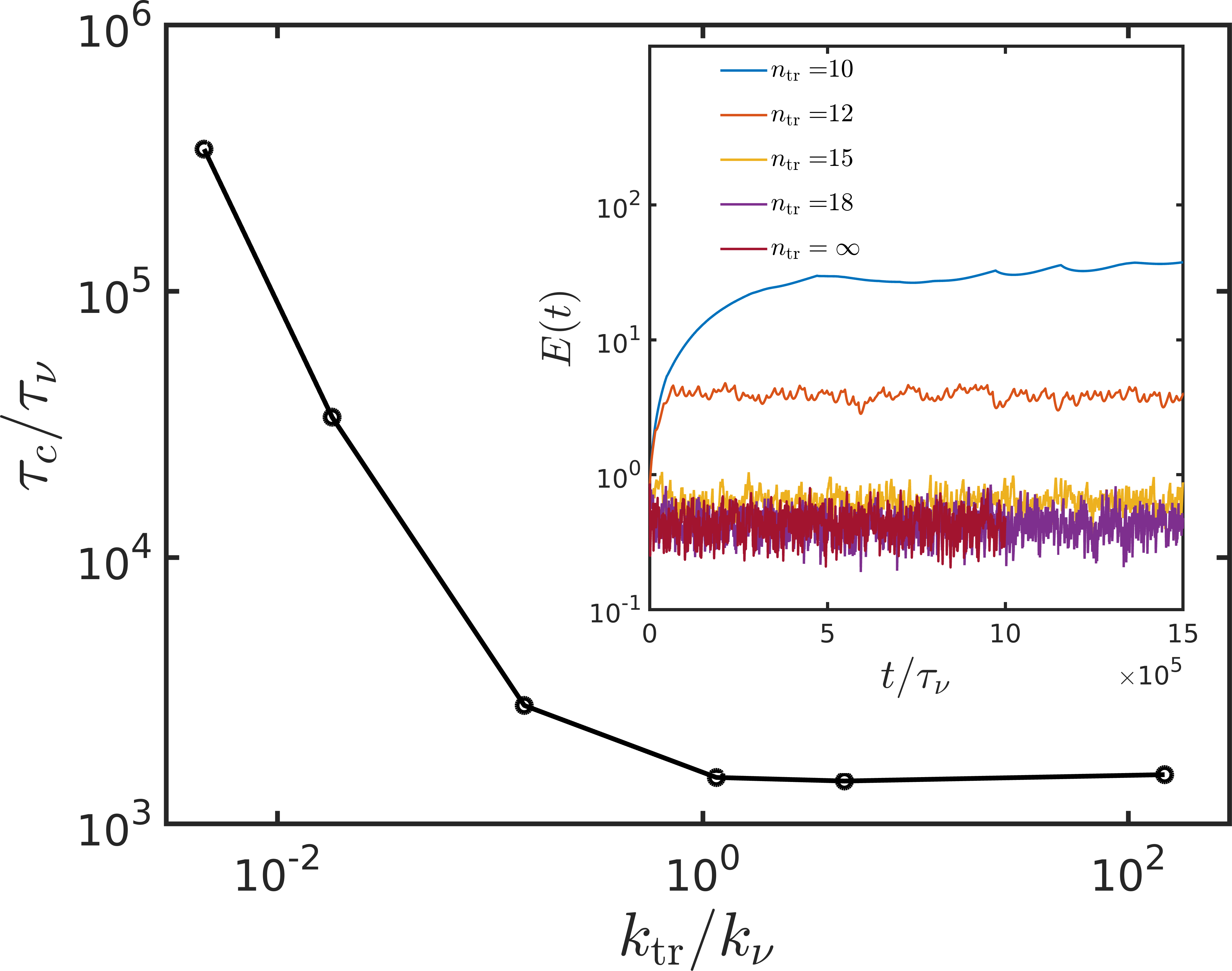}

    \caption{Plot of the energy-autocorrelation decay time $\tau_c$, scaled by the Kolmogorov time $\tau_\nu=\sqrt{\nu/\epsilon_{f}}$, as a function of $\ktr/k_{\nu}$. Inset: the time series  of the total energy for different values of $\ntr$ (runs A0, A3, A4, A5, and A6).}
    \label{fig:et_c1}
\end{figure}
\begin{figure*}[!]
      \resizebox{\linewidth}{!}{ 
    \includegraphics[width=\linewidth]{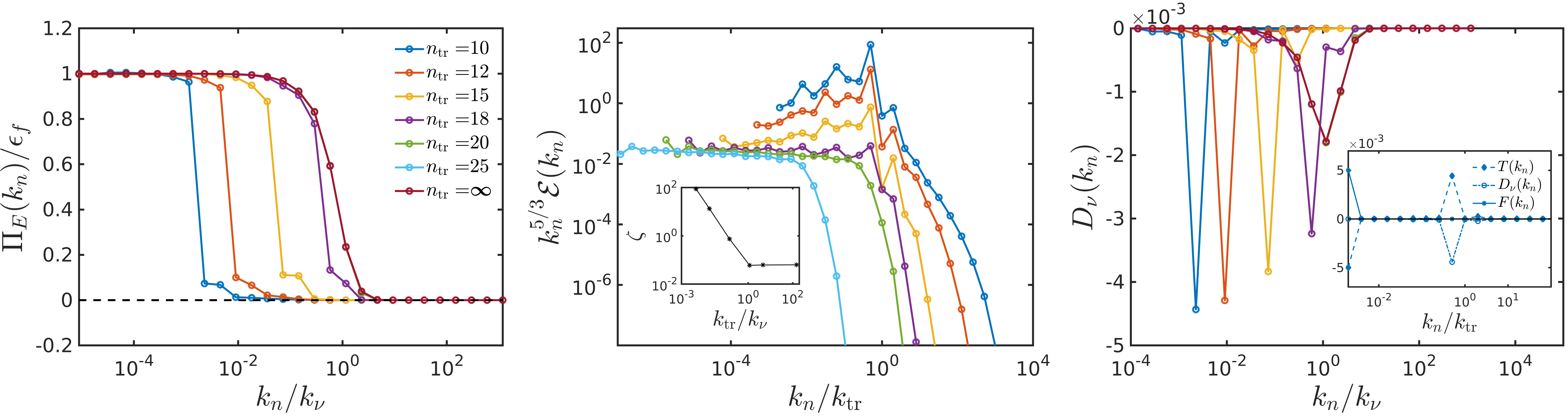} 
    	\put(-510,115){\rm{(a)}}
		\put(-335,115){\rm {(b)}}
		\put(-160,115){\rm {(c)}}
		}
    \caption{(a) Log-linear plot of the scaled energy flux versus $\kn/k_{\nu}$ for different values of $\ntr$ (runs A0 to A6); the plots for $\ntr=20, 25$ are indistinguishable from the $\ntr=\infty$ curve. \DVred{The Kolmogorov wavenumber corresponds to $n$ lying between 17 and 18} (b) Log-log plots of the compensated energy spectra versus $\kn/\ktr$ for the values of $\ntr$ in (a)
    ($\ktr = k_0 \lambda^{\ntr}$); inset: the value of the peak $\zeta$ of the compensated spectrum versus $\ktr/k_{\nu}$. (c) Log-linear plot of $D_{\nu}(\kn)$ versus $\kn/k_{\nu}$ for different values of $\ntr$ (runs A); the plots for $\ntr=20, 25$ are indistinguishable from the $\ntr=\infty$ curve; inset: log-linear plots of $T(\kn)$, $D_{\nu}(\kn)$, and $F(\kn)$ versus $\kn/k_{\nu}$ for  
 the representative value $\ntr=10$. The color coding for $\ntr$ is the same in (a)-(c).}
    \label{fig:efx_c1}
\end{figure*}

\DVblue{Now consider} $1 < \ktr/k_f \ll \infty $. If $\ktr\gtrsim k_{\nu}$, $E(t)$ does not show significant deviations from the limiting case 
$\ktr/k_f\to\infty$ (inset of Fig.~\ref{fig:et_c1}).
However, when $\ktr<k_{\nu}$ we find that, with decreasing $\ktr$ (and fixed $k_{f}$), $E(t)$ takes longer to reach the stationary state, and its stationary value increases (inset of Fig.~\ref{fig:et_c1}). To characterize the temporal energy fluctuations $E'(t) = E(t) - \langle E(t) \rangle$ for different $\ktr$, we calculate
the time scale $\tau_{c}$ from the exponential decay of the autocorrelation function $C(\tau)=\langle E'(t+\tau)E'(t) \rangle/\langle E'^{2}(t)\rangle$. Figure~\ref{fig:et_c1} shows that $\tau_c$ is nearly independent of $\ktr$, for $\ktr\geq k_{\nu}$, but it grows rapidly as $\ktr$ is decreased below $k_\nu$. These trends are reminiscent of the formation of large-scale condensates in 2D turbulent flows~\cite{be12,chan2012dynamics}, where, in the absence of friction, condensate formation is associated with very long saturation times and strong deviations of the energy spectrum from its inertial-range scaling (compare Fig. 2 of Ref.~\cite{chan2012dynamics} with our Fig.~\ref{fig:et_c1}). Therefore, we examine the dependence of energy spectra and flux on $\ktr$. In Fig.~\ref{fig:efx_c1}(a), we plot $\Pi_{E}(\kn)$ [Eq.\eqref{Eflux}] for different values of $\ktr$. 
As long as $\ktr>k_{\nu}$, this flux is indistinguishable from 
that of the $\ktr/k_f\to\infty$ case. However,
if $\ktr<k_{\nu}$, $\Pi_E(\kn)\simeq\epsilon_{f}$ only for $k_{f}<\kn<\ktr$, and it vanishes rapidly beyond $\ktr$.
Thus, the direct energy cascade persists up until $\ktr$, but is then arrested by the 2D-like dynamics for $\kn > \ktr$. The consequence of this arrest is a sharp build-up of energy around $\ktr$, as seen in Fig.~\ref{fig:efx_c1}(b), where we plot the compensated energy spectra $\kn^{5/3}\mathcal{E}(\kn)$ 
versus $\kn/\ktr$, for different values of $\ktr$.
Energy starts accumulating around $\ktr$ for $\ktr<k_\nu$; this accumulation increases as $\ktr$ approaches $k_f$. We also remark that the suppression of high-frequency fluctuations in $E(t)$,  with decreasing $\ktr$, is associated with the arrest of the energy cascade at $\kn\approx\ktr$.

\DVblue{Despite the} build up of energy at a scale smaller than $k_{\nu}$, our model reaches a statistically stationary state, albeit at times
that increase as $\ktr$ decreases. To understand this intriguing behavior, we examine the energy budget in the statistically stationary state:
\bel{ebalance}
T(\kn)+D_{\mu}(\kn)+D_{\nu}(\kn)+F(\kn)=0.
\ee
\DVblue{Here,} $T(\kn)=\langle \Re\{\mathrm{i}\nln\un^{*}\}\rangle,~~F(\kn)=\langle \Re\{\fn\un^{*}\}\rangle,~~ 
D_{\nu}(\kn)=\langle2\nu\kn^{2}\vert\un\vert^{2}\rangle,~{\rm{and}}~D_{\mu}(\kn)=\langle  2\mu\kn^{-2}\vert\un\vert^{2}\rangle$
are the nonlinear, forcing, viscous, and friction contributions, respectively. \DVblue{We plot these quantities} in the inset of Fig.~\ref{fig:efx_c1}(c) for $\ntr=10$.
Since friction is absent, the forcing term is balanced by the transfer term at $\kn=k_{f}$, and, in the cascade range $\kn<\ktr$, the contribution from the transfer term is negligible, 
 so the statistical properties are like those in the pure 3D direct cascade
 ($\ktr/k_f\to\infty$). Deviations from this 3D cascade arise when we
 account for the dissipation term. The maximum of $\vert D_{\nu}(\kn)\vert$ shifts from $\kn\simeq k_{\nu}$ to $\kn\simeq\ktr$, thus compensating for the accumulation of energy at the same wavenumbers. \DVblue{This is clearly shown} in the inset of Fig.~\ref{fig:efx_c1}(c), where $T(\kn)$ is balanced by $D_{\nu}(\kn)$ at $\kn\simeq \ktr$. Moreover, as $\ktr$ decreases, the maximum of $\vert D_{\nu}(\kn)\vert$ shifts to smaller 
 values of $\kn$ and its magnitude increases to compensate for the stronger build-up of energy [Fig.~\ref{fig:efx_c1}(c)]. 

\begin{figure}[!]  
\includegraphics[width=0.8\columnwidth]{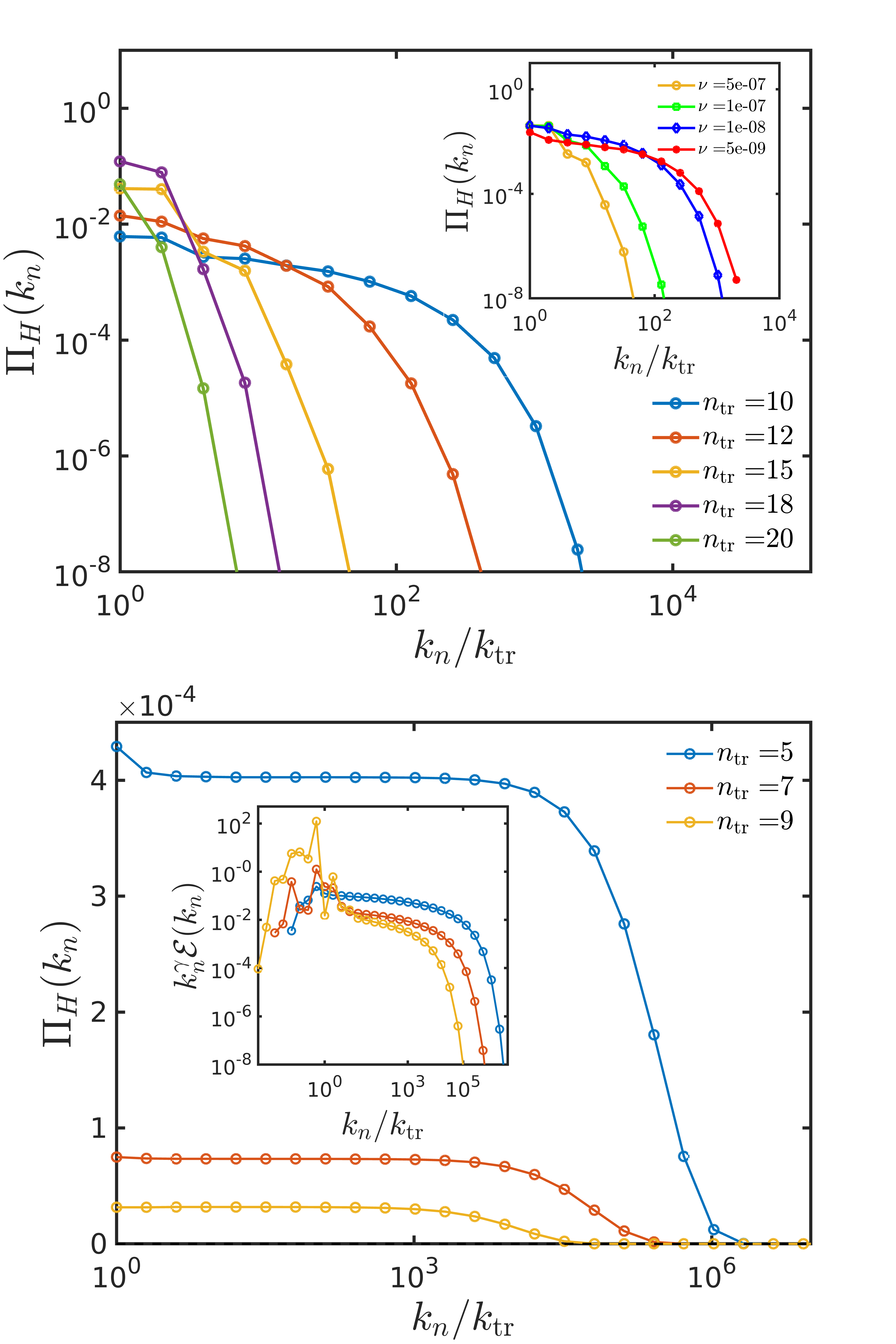}
           \put(-200,270){\rm{(a)}}
              \put(-200,135){\rm{(b)}}

    \caption{Log-linear plots of (a) $\Pi_{H}(\kn)$ versus $\kn/\ktr$ for runs A1-A6  and, in the inset, for runs A6, B1, B2 and B3, and (b) for runs C1-C3. 
    Inset of (b): log-log plots of compensated energy spectra versus $\kn/\ktr$. }
    \label{fig:hfx_c1}
\end{figure}

In the range $\kn>\ktr$, the coefficients in Eq.~\eqref{model} lead to 2D-like dynamics; hence, $H(t)$ is both positive definite and conserved locally.
By analogy with the inverse-cascade regime in 2D fluid turbulence, we expect that the energy that accumulates at $\kn\simeq\ktr$ acts as a source for the direct cascade of $H$ in the range $\ktr\ll \kn\ll k_\nu$. To confirm this, we plot,
in Fig.~\ref{fig:hfx_c1}(a), the flux of $H$:
\bel{fluxH}
\Pi_{H}(\kn)=\bigg\langle\sum^{N}_{j=n}\Re\big\{\mathrm{i}\kn^{\beta}\Phi_{j}u^{*}_{j}\big\}\bigg\rangle
\ee
with
$\beta=\log_{\lambda}(a/c)$. 
Clearly, as $\ktr$ is decreased, the flux increases and tends to flatten for $\ktr\ll \kn\ll k_\nu$, suggesting a direct cascade of $H$. However, the lack of significant separation between $\ktr$ and $k_{\nu}$ makes it difficult to identify a range where $\Pi_{H}(\kn)$ remains constant. 
In the inset of Fig.~\ref{fig:hfx_c1}(a), we plot $\Pi_{H}(\kn)$ for fixed $\ktr$ and different values of $\nu$ to observe indeed that, for small viscosities, $\Pi_{H}(\kn)$ tends to flatten for $\ktr\ll \kn\ll k_\nu$. 
As further confirmation of the direct cascade of $H$, we show in Fig.~\ref{fig:hfx_c1}(b) that, by 
moving $\ktr$ close to $k_1$, we achieve a large range of constant $\Pi_{H}(\kn)$. \DVblue{Moreover, in such range,} 
$\mathcal{E}(k_n)\sim k_n^{-\gamma}$ with $\gamma=2[1+\beta]/3+1$ [inset of Fig.~\ref{fig:hfx_c1}(b)], as is expected in the direct-cascade regime of $H$ \cite{gilbert02}.

\begin{figure}[!]
       \includegraphics[width=0.8\columnwidth]{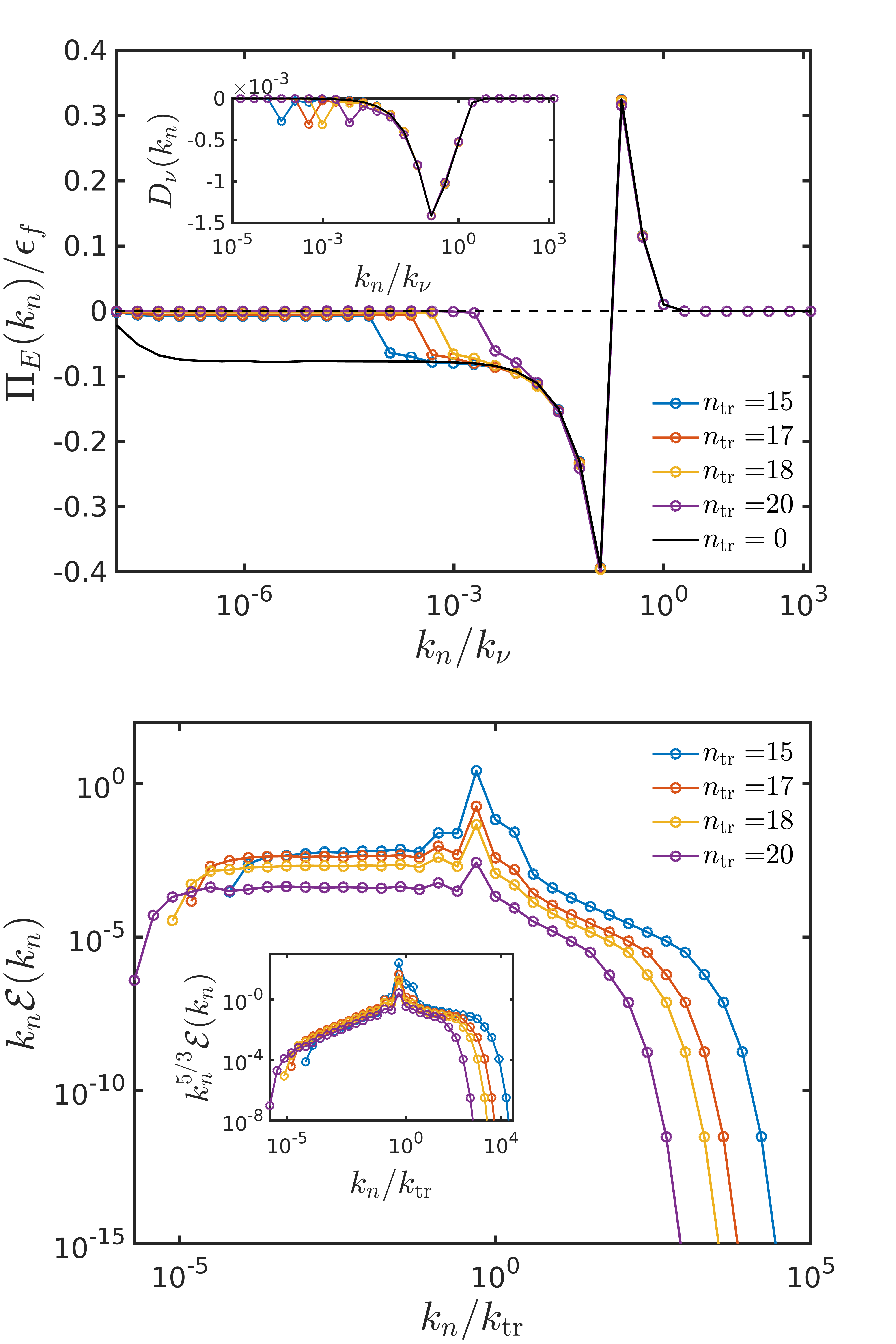}
              \put(-200,270){\rm{(a)}}
              \put(-200,135){\rm{(b)}}
     \caption{Plots for runs D1-D5: (a) Log-log plot of the compensated spectra $\kn\mathcal{E}(\kn)$ versus $\kn/\ktr$ and in the inset the compensated spectra $\kn^{5/3}\mathcal{E}(\kn)$ versus $\kn/\ktr$. (b) Log-linear plots of $\Pi(\kn)$ versus $\kn/k_{\nu}$ and in the inset $D_{\nu}(k_{n})$ versus $\kn/k_{\nu}$. \DVred{The Kolmogorov wavenumber corresponds to $n$ lying between 27 and 28}
     }
    \label{fig:efx_c2}
\end{figure}

\paragraph{(b) Large-wavenumber forcing.}
We now address the case $k_{f} > \ktr$ (Table~\ref{tab:parameters}, runs~D;
$\mu\neq0$ helps the system to reach a statistically stationary state \DVred{but does not contribute to the formation of the condensate \footnote{We have repeated runs D2-D5 with $\mu=0$, and the conclusions remain the same}}).
For  $\ktr=0$ and with our choice of parameters,
$\mathcal{E}(k_n) \sim k_n^{-5/3},$ in the range between the friction-dominated wavenumbers and $k_f$ \cite{gilbert02}.
In this range, $\Pi_E(k_n) < 0$  remains constant and  equals the rate of hyper-friction energy dissipation $\epsilon_\mu=\sum^{N}_{n=1}D_{\mu}(\kn)$. 
We now consider $k_{f} > \ktr > 0$. Small values of $\ktr$ have a negligible effect
on energy spectra and fluxes, because at small $\kn$ the inverse energy cascade is already stopped by hyper-friction; so we focus on intermediate values of $\ktr$ between $k_1$ and $k_f$. \DVblue{For such values of $\ktr$,} the energy flux  [Fig.~\ref{fig:efx_c2}(a)] indicates that the inverse energy cascade is arrested at $\kn\simeq\ktr$ and it is accompanied by energy build-up \DVred{(i.e.~the formation of an intermediate-scale condensate)} around $\ktr$ [Fig.~\ref{fig:efx_c2}(b)]. 

We see the following two scaling forms
on each side of $\ktr$: For $k_1\ll\kn\ll k_{tr}$, $\mathcal{E}(\kn) \sim 
\kn^{-1}$, which
indicates equipartition [Fig.~\ref{fig:efx_c2}(b)]; and 
for $\ktr\ll\kn\ll k_f$, $\mathcal{E}(\kn) \sim \kn^{-5/3}$, as we expect in the range of the inverse energy cascade [inset of Fig.~\ref{fig:efx_c2}(b)]. Clearly, the latter range decreases as $\ktr$ approaches $k_f$. 
The energy transfer that leads to a statistically stationary state is similar to that observed in case $(a)$: the accumulation of energy at scales comparable to $\ktr$ is compensated by an increased viscous dissipation at similar scales. Indeed, in Fig.~\ref{fig:efx_c2}(a) we see, together with a peak of dissipation at
$k_f$, a second peak at $\ktr$ \footnote{We have checked that the contribution of hyper-friction to the formation of the energy condensate is negligible.}.
\DVred{Before concluding, we comment on the aforementioned regime of energy equipartition.
In shell models, with constant 3D-like coefficients, the wavenumbers $k_{n} \ll k_{f}$ are isolated from the effects of forcing and viscous damping, and the energy flux to these small wavenumbers is zero~\cite{ccv24}. Therefore, energy in these modes is expected to equilibrate \cite{frisch1995,ccv24}. Analogous arguments lead to energy equipartition for $\kn\ll\ktr$ in our model: the condensate generated around $\ktr$ by the large-$\kn$ inverse cascade indeed acts as a source of energy for the 3D-like low-$\kn$ modes. 
It remains to be understood whether or not such a regime is expected to persist in real turbulent flows.
Energy equipartition at length scales larger than the forcing length scale has been reported in both numerical and experimental investigations~\cite{dallas2015statistical,alexakis2019thermal,gorce2022statistical}. However, a recent study~\cite{Ding_Xie_Wang_2024} has shown that, in 3D homogeneous and isotropic turbulence, the statistics of the large scales is not Gaussian. This has been attributed to a strong coupling between the small and large scales, which may not be captured by shell models. Furthermore,
numerical simulations of 3D turbulence have reported significant deviations from the equilibrium spectrum for certain forcing choices~\cite{alexakis2019thermal,schekochihin2023thermal}. Therefore, we expect that spatial and temporal features of the intermediate-scale condensate, specific to the hydrodynamical PDEs, are detrimental to the presence of equipartition at large length scales. 
}

\paragraph{Conclusions.}~Our study \KVK{identifies} a general energy-transfer mechanism for the non-dissipative arrest of energy cascades (inverse or direct) when 3D turbulent dynamics, at small $k$, coexists with 2D turbulent dynamics, at large $k$. The shell-model approach we employ allows us to \KVK{resolve} a large range of wavenumbers; this is crucial for uncovering the subtle interplay between the nonlinear and viscous terms. Specifically, we find that, when $\ktr/k_{f}>1$, the direct energy cascade for $\kn<\ktr$ is arrested by the 2D-like dynamics at $\kn>\ktr$. In contrast, when $\ktr/k_{f}<1$, the inverse energy cascade for $\kn>\ktr$ is arrested by the 3D-like dynamics at $\kn<\ktr$. In both cases, the arrest close to $\ktr$, results in energy accumulation around $\ktr$. In a spatially extended system, such an accumulation of energy would lead to the emergence of spatial structures, or condensates, of size $\ktr^{-1}$. A statistically stationary state stems from an increased viscous dissipation, at wavenumbers close to $\ktr$, which compensates for the energy accumulation. This is reminiscent of condensates in 2D turbulence. Furthermore, we show that, when $\ktr/k_{f}>1$, the energy that accumulates near to $\ktr$ generates a direct cascade of generalized enstrophy for $\kn>\ktr$, whereas, when $\ktr/k_{f}<1$, the modes with $\kn<\ktr$ are in statistical equilibrium.

\KVK{The details of the transitional scale will vary with different models, but the energy-transfer mechanisms we have identified are general and apply to any turbulent system with 2D-like large-$k$ dynamics and 3D-like small-$k$ dynamics. Thus, our results will stimulate the study of new physical systems with cascade arrests leading to intermediate-scale condensates.}

\begin{acknowledgments}
This work was supported in part by the French government through the UCA$^{\rm JEDI}$ Investments in the Future project managed by the National Research Agency (ANR), with reference number ANR-15-IDEX-01, and by the International Centre for Theoretical Sciences (ICTS) for the online program - Turbulence: Problems at the Interface of Mathematics and Physics (code: ICTS/TPIMP2020/12).
KVK thanks Laboratoire J.~A.~Dieudonn\'e of Universit\'e C\^ote d'Azur for hospitality. RP and KVK acknowledge support from SERB and NSM (India) and thank
SERC (IISc) for computational resources.
KVK and DV are thankful to Siddhartha Mukherjee for useful discussions.
DV acknowledges support from the International Centre for Theoretical Sciences (ICTS-TIFR), Bangalore, India.
\end{acknowledgments}
\bibliography{references}

\end{document}